\documentstyle[11pt,aaspp4]{article}

\slugcomment{Accepted for publication in ApJ}

\lefthead{}
\righthead{Broad-line region search in NGC 1068}

\begin{document}

\title{A search for broad infrared recombination lines in NGC\,1068
\footnote{Based on observations with ISO, an
   ESA project with instruments funded by ESA Member States (especially the PI
   countries: France, Germany, the Netherlands and the United Kingdom) with the
   participation of ISAS and NASA.}
   }

\author{D.~Lutz\footnote{Max-Planck-Institut f\"ur extraterrestrische Physik,
   Postfach 1603, 85740 Garching, Germany},
   R.~Genzel$^2$,
   E.~Sturm$^2$,
   L.~Tacconi$^2$,
   E.~Wieprecht$^2$,
   T.~Alexander\footnote{Institute for Advanced Study, Olden Lane, 
         Princeton, NJ 08540, USA},
   H.~Netzer\footnote{School of Physics and Astronomy and Wise Observatory,
      Raymond and Beverly Sackler Faculty of Exact Sciences, Tel Aviv 
      University, Ramat Aviv, Tel Aviv 69978, Israel},
   A.~Sternberg$^4$,
   A.F.M.~Moorwood\footnote{European Southern Observatory,
     Karl-Schwarzschild-Stra\ss\/e 2, 85748 Garching, Germany},
   R.A.E.~Fosbury$^5$,
   K.~Fricke\footnote{Universit\"atssternwarte G\"ottingen, 
        Geismarlandstra\ss\/e 11, 37083 G\"ottingen, Germany},
   S.J.~Wagner\footnote{Landessternwarte, K\"onigstuhl, 69117 Heidelberg, 
       Germany},
   A.~Quirrenbach\footnote{Center for Astrophysics and Space Sciences,
             University of California, San Diego, Mail Code 0424,
             La Jolla, CA 92093-0424,USA},
   H.~Awaki\footnote{Department of Physics, Faculty of Science, Kyoto 
         University, Sakyo-ku, Kyoto, 606-01, Japan},
   K.Y.~Lo\footnote{Institute of Astronomy and Astrophysics, Academia Sinica, 
    Taipei 115, Taiwan, ROC}
   }
\begin{abstract}

We report infrared spectroscopy of the prototypical Seyfert~2 galaxy NGC 1068,
aiming at detection of broad components of hydrogen recombination lines
that originate in the obscured broad-line region. Using the Short Wavelength 
Spectrometer on board the Infrared Space Observatory, we have observed for 
the first time the regions of Brackett\,$\beta$ 2.626$\mu$m and 
Pfund\,$\alpha$ 7.460$\mu$m, and
present improved data for Brackett\,$\alpha$ 4.052$\mu$m. No significant broad
components are detected, implying an equivalent visual extinction
to the broad-line region of at least 50 magnitudes and an obscuring
column density of at least $10^{23}$cm$^{-2}$. While consistent
with a highly obscured broad-line region, as required by the classical unified
scenario, these limits are not yet 
significant enough to discriminate strongly between different torus models or 
to constrain properties of the gas causing the very large X-ray obscuration.
We discuss the systematic limitations of infrared broad-line region searches
and suggest that Brackett\,$\alpha$ may often be the most favorable transition
for future searches.

\end{abstract}


\section{Introduction}

The prototypical Seyfert~2 galaxy NGC\,1068 has played a key role in the
development of `unified' models in which Seyfert~2 galaxies host a 
Seyfert~1 -- like broad-line region (BLR). These models assume that the Seyfert~2 BLR 
is obscured toward our line of sight by a dusty torus, and
have been highly successful in explaining several aspects of 
Seyfert galaxies. The key observational evidence supporting such models
came from the spectropolarimetric dectection of broad
hydrogen recombination lines, first in NGC\,1068 (\cite{antonucci85}) and later
in a number of other Seyfert~2s. Direct detection of the obscured BLR may 
be, in principle, possible at near- and mid-infrared wavelengths and can
provide another route to studies of Seyfert unification, independent
of the uncertain scattering efficiency entering the quantitative analysis of 
most spectropolarimetric data. In addition, if good
detections or limits could be obtained for a significant sample of Seyfert~2s
it would be possible to place constraints on column densities and geometries
of the putative tori, by studying detection rates and columns as a function
of orientation. 

Successes have been reported in detecting broad components of
near-infrared recombination lines in several type 2 Seyferts (e.g. \cite{rix90};
\cite{blanco90}; \cite{goodrich94}; \cite{ruiz94}; \cite{veilleux97}). 
However, the observed FWHM line widths
(1000 to 3000 km/s) are typically narrower than those of most classical
broad-line regions. In NGC 5506, for example, the broadish component may 
trace an obscured narrow-line region (NLR)
component rather than a true dense BLR (\cite{goodrich94}). In addition, 
column densities of a few times 
10$^{22}$cm$^{-2}$ as accessible to near-infrared spectroscopy fall short of
the column densities expected from pc-scale torus scenarios 
(\cite{krolik88}), and barely reach column densities directly inferred from
mm interferometric observations of near-nuclear gas in NGC\,1068
(e.g. \cite{tacconi94}). 
Near-IR broad-line detections apparently pick a biased subsample:
some of the detections (NGC 2992, NGC 5506, A0945-30), in fact, 
refer to narrow-line X-ray galaxies, in which the detection
of strong X-rays suggests a significantly lower obscuring column than in
most classical Seyfert~2s. Some of them had been previously suggested to be 
intermediate Seyferts on the basis of possible broad lines in their 
optical spectra.

X-ray observations are, by far, the most sensitive large column
density indicators. The measured obscuring columns in Seyfert 2 galaxies 
range from ${\rm few} \times 10^{20}$ cm$^{-2}$ to $> 10^{25}$ cm$^{-2}$
(e.g. \cite{turner97}; \cite{bassani99}) suggesting that near-infrared
lines may be usually fully absorbed. However, the relation between 
X-ray column and infrared obscuration may depend on properties of
the obscuring material.
There is, hence, a strong rationale for trying to detect longer wavelength 
recombination 
lines that are capable of penetrating a larger column, and to extend searches 
to more
difficult but more rewarding X-ray quiet targets. In the following,
we report on upper limits for broad 2.62$\mu$m Brackett\,$\beta$, 
4.05$\mu$m Brackett\,$\alpha$ and 
7.46$\mu$m Pfund\,$\alpha$ emission from the nuclear region of NGC\,1068
which has one of the largest X-ray columns.

The major observational difficulties in attempts to detect broad mid-infrared 
recombination lines are (1) the rapid fall of intrinsic recombination line 
fluxes towards higher series, i.e. longer
wavelengths; (2) the active galactic nucleus (AGN) dust continuum rising 
steeply towards longer
wavelengths, leading to small line-to-continuum ratios vulnerable to
systematic effects; and (3) the shape of the mid-infrared extinction
curve which determines the wavelengths capable of penetrating the highest
obscuring column. For a galactic type extinction law (e.g. \cite{draine89}), the 
Pfund\,$\alpha$ 7.46$\mu$m line might be a `sweet spot' in being 
intrinsically still quite bright but suffering little obscuration in the deep
extinction minimum near 7$\mu$m before the onset of silicate absorption.
The same low extinction, able to penetrate columns of 
$\gtrsim\/10^{23}$cm$^{-3}$, is only found again at very long wavelengths
$\gtrsim$30$\mu$m where the recombination lines are intrinsically very faint
and exceedingly difficult to detect against the strong dust continuum.

\section{Observations and limits on emission from the hidden BLR}

We have used the Short Wavelength Spectrometer SWS (\cite{degraauw96}) on 
board the Infrared Space Observatory ISO (\cite{kessler96}) to search for
broad hydrogen recombination lines in NGC\,1068.
The observations were targeted at Brackett\,$\beta$ 2.62$\mu$m, 
Brackett\,$\alpha$ 4.05$\mu$m and Pfund\,$\alpha$ 7.46$\mu$m, using the
SWS06 mode to cover ranges of $\sim$\,6000km/s (Brackett\,$\alpha$ which is close
to an SWS `band edge', see below) to $\sim$\,14000km/s (Pfund\,$\alpha$) at
a good signal-to-noise ratio. The 14\arcsec$\times$20\arcsec\/ aperture
was oriented approximately north-south and centered on the nucleus, thus
covering any hidden BLR, the NLR and part of the extended emission line region.
Humphreys\,$\alpha$ 
(n=7-6) 12.37$\mu$m was observed as well but does not provide a useful
measurement since the SWS relative spectral response in this wavelength
range is less well defined due to a very complex fringing pattern.  
We have analyzed the data using the SWS Interactive Analysis (IA) system
(\cite{lahuis98}; \cite{wieprecht98}) and calibration files as of July 1998.
A more detailed discussion of the data analysis is given by Lutz
et al. (2000) together with a presentation of our complete SWS dataset on
NGC\,1068.

Broad emission is not detected significantly in any of the three lines. 
We will now discuss the upper limits
on broad components of the various lines (Table~\ref{tab:blrlimits}), 
adopting $\sim$3000km/s as the
FHWM of the broad emission. This is the FHWM measured
for NGC\,1068 in polarized H$\beta$ (\cite{miller91}, \cite{inglis95})
from regions where linewidth-preserving dust scattering dominates the 
scattering process. Particular attention is given to the effects of
relative spectral response calibration and of broad emission features 
from complex molecules in the ISM on the determination of the
underlying continuum. These are limiting factors in the search for the BLR
in the strong near- to mid-infrared continuum of NGC\,1068.

At Brackett\,$\beta$, NLR emission is clearly seen (Figure~\ref{fig:brb}). 
Lutz et al. (2000) find NLR fine structure lines over a wide range
of conditions reasonably well fitted by a profile with two gaussian
components, having FWHM 333 and 1246\,km/s, peak ratio narrow/wide
1.34, and the wider component blueshifted by 100km/s. This is consistent
with the  Brackett\,$\beta$ profile within the (considerable) noise.
Brackett\,$\beta$ is blended with the (1-0) O(2) ro-vibrational
transition of molecular hydrogen. Based on other observed
H$_2$ lines (\cite{lutz00}), this transition is, however, 
estimated to be a minor
contributor ($\lesssim$10\%) to the flux for the case of NGC\,1068. 
Our limit on broad (FWHM 3000 km/s)
emission is mainly set by noise despite an integration time of 
$\approx$3 hours. The line is observed close to the edge of one of the SWS
`AOT bands' (\cite{degraauw96}) . This implies increased uncertainties in 
the relative spectral
response calibration compared to the band centers, but there is no manifest 
effect on the Brackett\,$\beta$ data, since the variation
of sensitivity with wavelength is slow for this particular band limit. Some
of the large scale curvature in the Brackett\,$\beta$ spectrum may be due to
this effect. The curvature almost certainly does not suggest an extremely 
broad BLR component and would be inconsistent with the spectropolarimetric
evidence on the width of BLR lines. We note 
that the S/N of the data approaches the limit given by the (average) noise 
in the relative spectral response function of the twelve detectors at this
wavelength.

For Brackett\,$\alpha$, the relative spectral response changes 
by a factor of almost two
over the short range covered by the line (Figure~\ref{fig:bra}) which is at the
long wavelength end of the SWS AOT band 1E. Under these conditions,
systematic effects are very important in addition to the (low)
detector noise. There may 
be residual systematic uncertainties in the spline smoothing process used for
in-orbit calibration of the SWS relative spectral response
(B. Vandenbussche, priv. comm.). Also, small inaccuracies of dark current
subtraction can imprint a residue of the relative spectral response function
on the final spectrum. The wavelength 
range above 4.08$\mu$m was observed as well, but is not shown since it does 
not contribute to the Brackett\,$\alpha$ profile because of much higher 
detector noise.
When overplotting the Brackett\,$\alpha$ profile
with the NLR profile which fits the observed fine structure lines quite well
(Fig.~\ref{fig:bra}),
or attempting fits of two gaussians on a sloped continuum,
the wings of the Brackett\,$\alpha$ profile appear slightly wider than for
the typical NLR profile. We do not 
consider this a detection of a broad component, however, because the excess 
would be $\lesssim$1\% of the continuum and not fully robust with respect to
the mentioned spectral response calibration uncertainties and to location
of the continuum.
In Table~\ref{tab:blrlimits} we quote a conservative upper limit for a broad 
component, estimated on the basis of possible systematic problems.  
The nondetection of broad Brackett\,$\alpha$ strengthens previous 
limits on such a component (\cite{depoy87}; \cite{oliva90}).
  
Figure~\ref{fig:pfa} shows the result of an $\sim$1 hour integration on the
Pfund\,$\alpha$ region. In this high S/N spectrum we see the wing of the 
very strong [\ion{Ne}{6}] line as well as faint 
[\ion{Na}{3}] 7.318$\mu$m and a possible unidentified feature at rest wavelength
$\sim$7.555$\mu$m. Pfund\,$\alpha$ is not clearly detected, although there is a 
marginally significant maximum at the expected redshift (observed 
wavelength 7.489$\mu$m) 
and strength of NLR Pfund\,$\alpha$, as predicted from NLR Brackett\,$\alpha$,
case B ratios, and assuming low NLR extinction.
Again, we have set a conservative limit on a possible broad-line region 
contribution to Pfund\,$\alpha$. The main limitation of Pfund\,$\alpha$ BLR
searches is the continuum definition rather than the signal-to-noise ratio
of the line: 
not only are there the
interfering lines of [\ion{Ne}{6}] and [\ion{Na}{3}] but also, more importantly,
the effect of continuum features. This is better illustrated in 
Figure~\ref{fig:ppah} which shows a larger part of the full SWS spectrum
of NGC 1068 (\cite{lutz00})
around Pfund\,$\alpha$ both as observed, and with a tentative addition of 
faint starburst activity (taken from a scaled spectrum of M\,82) to the 
spectrum of NGC 1068 dominated by NLR dust. It is obvious
that the starburst-related emission features usually ascribed to polycyclic
aromatic hydrocarbons (PAHs) will induce curvature of the 
`continuum' near Pfund\,$\alpha$ since this is where the main `7.7$\mu$m' 
PAH feature starts to rise. This curvature will make detection of very broad
lines difficult if there is significant circumnuclear star formation. In 
addition, the shapes of the PAH features
are known to vary in detail with local conditions (e.g. \cite{verstraete96}; 
\cite{roelfsema96}). While they are likely destroyed very close to
the central AGN, regions further out may contribute to a spectrum like
that of NGC\,1068 with PAH shapes differing from those for a canonical
starburst, making accurate correction difficult. Similar continuum definition
problems may arise at a lower level even if there is no active circumnuclear
starburst, since PAH emission will be widespread also in the more quiescent 
disk of the Seyfert host (e.g. \cite{mattila99}).

\section{Discussion}

What are the implications of our upper limits on broad 
lines for the structure of the obscuring material? The derived 
obscuring column density
sensitively depends on the adopted intrinsic BLR line fluxes which
are usually not known accurately. For the favorable case of NGC\,1068, we 
are able to derive estimates (Table~\ref{tab:blrlimits}) on the basis
of the intrinsic broad H$\beta$ flux derived from spectropolarimetry
(\cite{miller91}; \cite{inglis95}), and of case B recombination line ratios.
Case B is not strictly adequate for BLR conditions since large optical depths
are expected in the Balmer and Paschen lines (e.g. \cite{netzer90} and 
references therein). However, except for the leading line in each series, 
the IR to optical line ratios are not expected to deviate much from this 
simple approximation and we have therefore adopted the calculations by
Storey \& Hummer (1995) as our best estimates. The ratios are not sensitive
to the adopted density and temperature and we have used the  
T$_e=10000$K, n$_e=10^6$cm$^{-3}$ values computed from their code. 
The ratios relative to H\,$\beta$ are therefore 0.0426 for Brackett\,$\beta$, 
0.0725 for Brackett\,$\alpha$, and 0.0224 for Pfund\,$\alpha$.

We adopt an intrinsic broad H$\beta$ flux of 
$\sim1\times\/10^{-17}$W\,cm$^{-2}$ on the basis of the reassuringly consistent
spectropolarimetric estimates of Miller, Goodrich, \& Matthews (1991) and
Inglis et al. (1995). Inglis et al. estimate an intrinsic broad H$\alpha$ 
flux of $4.3\times\/10^{-17}$W\,cm$^{-2}$, consistent with our value for a 
typical Balmer decrement observed in AGN ($\simeq 4$). 
The intrinsic broad H\,$\beta$ flux of $2.48\times\/10^{-18}$W\,cm$^{-2}$
quoted by Miller, Goodrich, \& Matthews (1991) has to be corrected
upwards for our purposes: it is derived from the {\em observed} total narrow 
[\ion{O}{3}] 5007\AA\/ flux  (\cite{shields75}) and the broad H$\beta$ / 
narrow [\ion{O}{3}] ratio in scattered light, taken from the `NE knot' where
the scattering properties are most homogeneous. The extinction to the 
narrow-line region of NGC\,1068 is, however, known to be significant at the
wavelength of [\ion{O}{3}] 5007\AA\/. We correct to our intrinsic broad 
H$\beta$ flux of $1\times\/10^{-17}$W\,cm$^{-2}$, using an 
[\ion{O}{3}] 5007\AA\/ correction factor of $\sim$4 consistent with 
the available extinction studies (\cite{neugebauer80}; \cite{koski78};
\cite{ward87}). 
The spectropolarimetric estimates for the intrinsic BLR flux implicitly
assume that the scatterers in the `NE knot' have a similarly unobscured
view of the BLR {\em and} NLR.

It is evident from Table~\ref{tab:blrlimits} that the limits measured
by ISO imply a significant obscuration of the BLR in the infrared.
Independent of detailed flux estimates, this can be explained by the 
following simple argument: intrinsically, the broad component of H$\beta$
will dominate the total H$\beta$ flux. This is typical for Seyferts 1
in general, but for NGC\,1068 also demonstrated directly by spectropolarimetry
(\cite{miller91}; \cite{inglis95}). Then, if an observation detects a NLR
recombination line at good S/N but fails to detect its broad component, the
additional BLR obscuration at that wavelength must (still) be significant.

A Brackett\,$\alpha$ obscuration
of $>$2.4\,mag (Table~\ref{tab:blrlimits}) corresponds to an equivalent
visual obscuration of A$_V\gtrsim$50\,mag, and an obscuring column density
of more than $\approx 10^{23}$cm$^{-2}$. These quantities depend
somewhat on the adopted extinction curve (cf. Table~\ref{tab:blrlimits}) and 
on conversion to column density (e.g. for normal ISM conditions and dust-to-gas
ratio
N$_H$=A$_V\times 1.79\times 10^{21}$cm$^{-2}$, Predehl \& Schmitt 1995). 
A$_V\gtrsim$50\,mag is a lower limit using
the most conservative assumption of a Galactic center extinction curve with 
relatively high extinction in the 4-8$\mu$m range. If
broad Brackett\,$\alpha$ were detected in our data, it could not
be obscured by significantly more than 10$^{23}$cm$^{-2}$. This is well
below the $\gtrsim$10$^{24}$cm$^{-2}$ columns needed to fit the X-ray properties
of Compton-thick Seyferts, which consequently are also the
columns used by standard `compact' torus models (\cite{krolik88}). 
For NGC\,1068, X-ray spectra clearly indicate an even higher
obscuring column (e.g. \cite{marshall93}; \cite{matt97}).
We note 
however that high columns derived from X-ray spectra do not enforce a priori 
unobservability of the mid-infrared transitions. The relation between
X-ray based column and infrared extinction may vary with gas-to-dust ratio 
or dust properties of the obscuring gas. Empirically, the relationship 
between broad-line region optical/infrared obscuration and
X-ray absorbing column may vary considerably from 
source to source (see e.g. cases discussed in \cite{granato97}; 
Maiolino et al., in prep). Detections or limits on infrared
broad lines could, hence, determine the properties of the putative torus in
a way that is independent from X-ray spectroscopy. They could also discriminate
among the different classes of torus models, in particular the
very high A$_V$ compact ones  (e.g \cite{krolik88}; \cite{pier92}) from lower
A$_V$ ones where much of the X-ray obscuration will occur in
dust-free gas (e.g. \cite{granato94}). At this point, the limit
of A$_V\gtrsim$50 approaches column densities predicted by some
large scale torus models (A$_V\sim 72$ for NGC\,1068, \cite{granato97}) but 
is not 
sufficient to discriminate strongly between the different scenarios.
It is however fully consistent with the presence of large molecular
column densities inferred from millimeter interferometry 
(e.g. \cite{tacconi94}).

Our observations help to shed light on strategies for future infrared BLR 
searches.
If the extinction in the 2-8$\mu$m range follows one of the classical curves
e.g. the $\lambda^{-1.75}$ power law of Draine (1989), then Pfund\,$\alpha$
would be the most favorable line for penetrating the highest columns. For 
such an
extinction curve in conjunction with observations limited by instrumental 
sensitivity, lower extinction compensates for the weaker intrinsic strength
of Pfund\,$\alpha$. There are, however, indications from ISO spectroscopy
that the extinction at Pfund\,$\alpha$ may be almost as large as at 
Brackett\,$\alpha$ (\cite{lutz96}, \cite{lutz97}). This value
was derived mainly for the line of sight towards our Galactic center, 
where local conditions could obviously be different from an AGN torus. 
Other disadvantages of Pfund\,$\alpha$ are the lower line-to-continuum
ratio (for NGC\,1068 the continuum rises by a factor of $\sim$2.5 from
4 to 7.5$\mu$m), the crowded spectral region with nearby fine structure 
lines, strong PAH emission features, and the possibility that absorption 
features similar to those seen towards the Galactic center (\cite{lutz96}) 
may be present depending on the physical conditions of the absorber. 
When using NGC\,1068 as a template for future searches for 
broad Pfund\,$\alpha$,
two peculiarities should be kept in mind. (1) At all near- and mid-infrared 
wavelengths, NGC\,1068 is known for its strong dust continuum and
resulting unusually low NLR line-to-continuum ratios. This should increase the 
prospects of BLR detection in other Seyfert~2s with more favorable 
line-to-continuum ratios. (2) On the other hand, the proximity of 
NGC\,1068 minimizes
contamination by PAH emission since most of the circumnuclear star formation
occurs outside a nuclear aperture, even if it is as large as that of ISO.
PAH contamination will be a problem for many other Seyfert~2s 
(\cite{clavel98}), even if somewhat
smaller apertures can be employed.
We hence suggest Brackett\,$\alpha$
to be a more promising target for infrared BLR searches, being brighter and
on a weaker continuum free of strong nearby emission and absorption features. 
Attempts to detect
broad Pfund\,$\alpha$ will definitely have to use as small as possible apertures
to minimize PAH dilution from circumnuclear star formation.

\acknowledgments
We thank Roberto Maiolino for discussions. SWS and the ISO Spectrometer
Data Center at MPE are supported by DLR (DARA) under grants 50 QI 8610 8 and
50 QI 9402 3. We acknowledge support by the German-Israeli Foundation.
Astronomical Research at Tel Aviv university is supported in part by the
Israeli Science Foundation.

%
%

\clearpage
\begin{table}
\caption{Limits on broad-line region emission}
\begin{tabular}{lrrrrr}
\tableline
\tableline
Line           &Wavelength &Observed Flux\tablenotemark{a}
                                        &Intrinsic flux\tablenotemark{b}
                                               &A$_V$\tablenotemark{c}
                                                     &A$_V$\tablenotemark{d}\\
               &\/$\mu$m   &$10^{-20}$\,W\,cm$^{-2}$
                                        &$10^{-20}$\,W\,cm$^{-2}$&mag&mag\\
\tableline
Brackett\,$\beta$ &2.626      &$<9$     &43    &$>$23&$>$23\\
Brackett\,$\alpha$&4.052      &$<8$     &73    &$>$68&$>$47\\
Pfund\,$\alpha$   &7.460      &$<13$    &23    &$>$52&$>$14\\  
\tableline
\end{tabular}
\label{tab:blrlimits}
\tablenotetext{a}{Upper limits for a broad component with FWHM 3000km/s.}
\tablenotetext{b}{Intrinsic broad-line flux estimates derived for 
comparison from the unobscured broad H$\beta$ flux estimate of 
$1\times\/10^{-17}$\,W\,cm$^{-2}$ (see text), no extinction, and case B 
recombination line ratios (\cite{storey95}).}
\tablenotetext{c}{Equivalent visual extinction to the broad-line region,
adopting $\rm A_{Br\beta}/A_V=0.075$, $\rm A_{Br\alpha}/A_V=0.035$,
$\rm A_{Pf\alpha}/A_V=0.012$ derived from the $\lambda^{-1.75}$ infrared
extinction curve of Draine (1989) and $\rm A_V=5.9\times E(J-K)$.}
\tablenotetext{d}{Equivalent visual extinction to the broad-line region,
adopting $\rm A_{Br\beta}/A_V=0.075$, $\rm A_{Br\alpha}/A_V=0.51$,
$\rm A_{Pf\alpha}/A_V=0.44$ representing the extinction curve towards the
center of our galaxy (\cite{lutz97})}
\end{table}

%
%

\clearpage

\figcaption[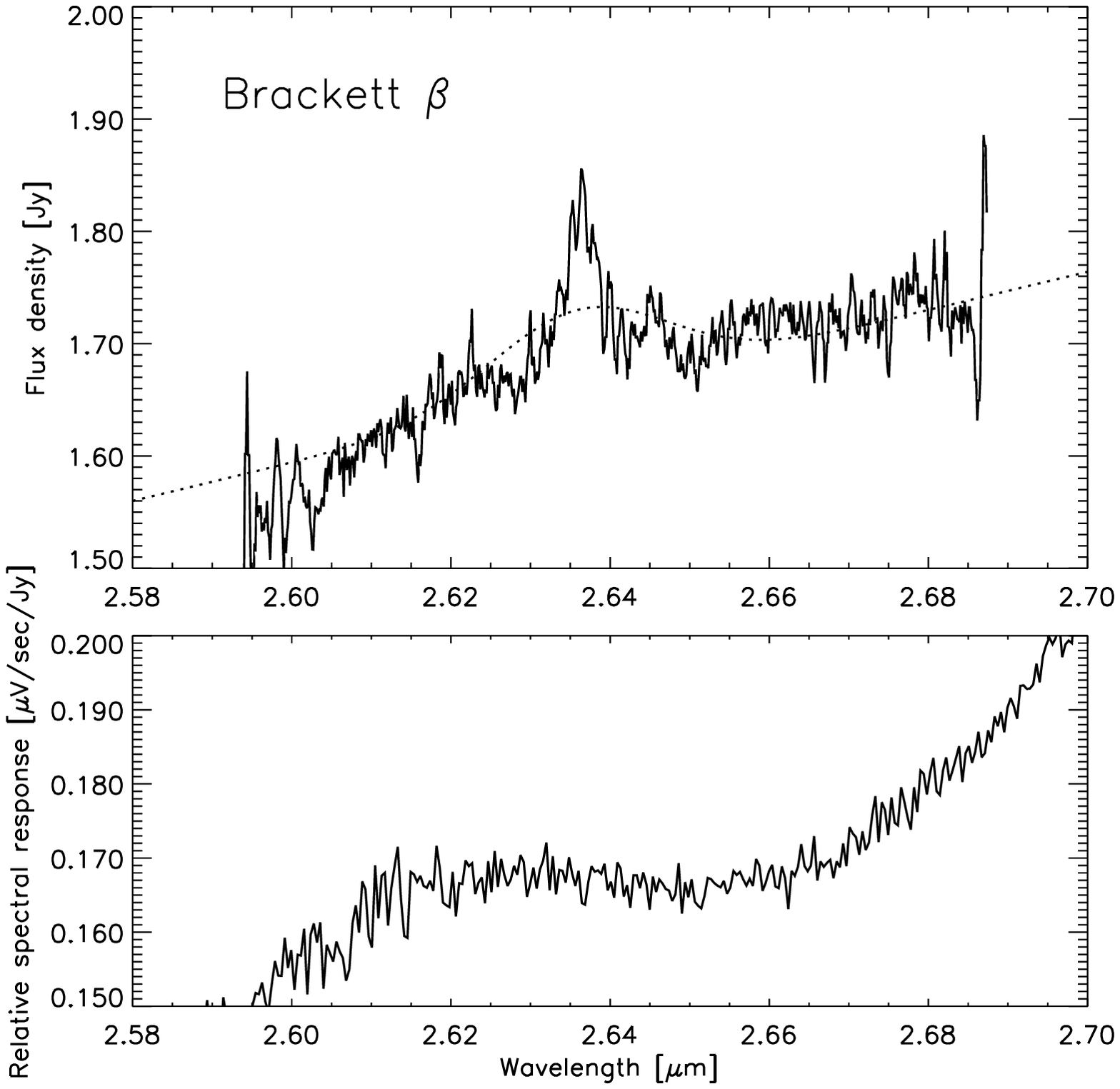]{  
Upper panel: Spectrum of the 2.626$\mu$m Brackett\,$\beta$ line in NGC\,1068. 
The dotted line indicates the suggested upper limit on a FWHM 3000km/s 
broad-line region component. Lower panel: The structure of the SWS relative 
spectral
response function is shown for one of the detectors used. There is only
moderate large scale structure despite the line being close to the edges of 
one of the SWS `AOT bands'. The high frequency modulation is detector noise 
from the
calibration used to derive the relative spectral response function.
\label{fig:brb}}

\figcaption[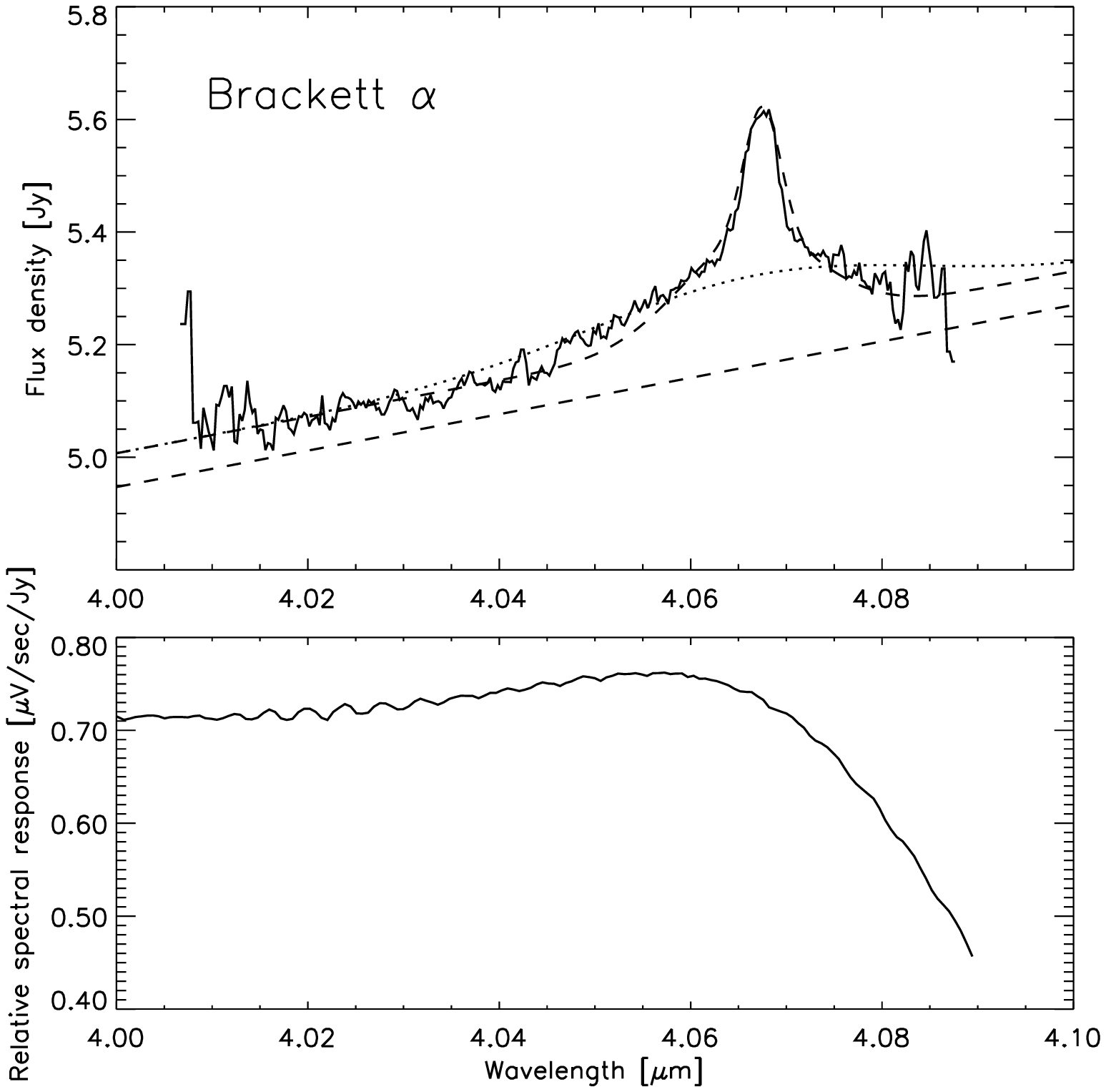]{
As in Figure~\ref{fig:brb}, for the 4.052$\mu$m Brackett\,$\alpha$ line. 
Again, the dotted line indicates the suggested upper 
limit on a FWHM 3000km/s broad-line region component. The lower dashed line 
represents the larger range continuum slope seen at 3.7 to 4.05$\mu$m in the 
full SWS spectrum (\cite{lutz00}), offset for clarity. The upper dashed 
line shows a continuum with the same slope plus a suitably scaled narrow line region
profile, as derived by Alexander et al. (2000) from mid-infrared fine structure
lines. Note the strong variation of the relative spectral response
over the short range covered by the line. Some fringing is seen in the
relative spectral response.
\label{fig:bra}}

\figcaption[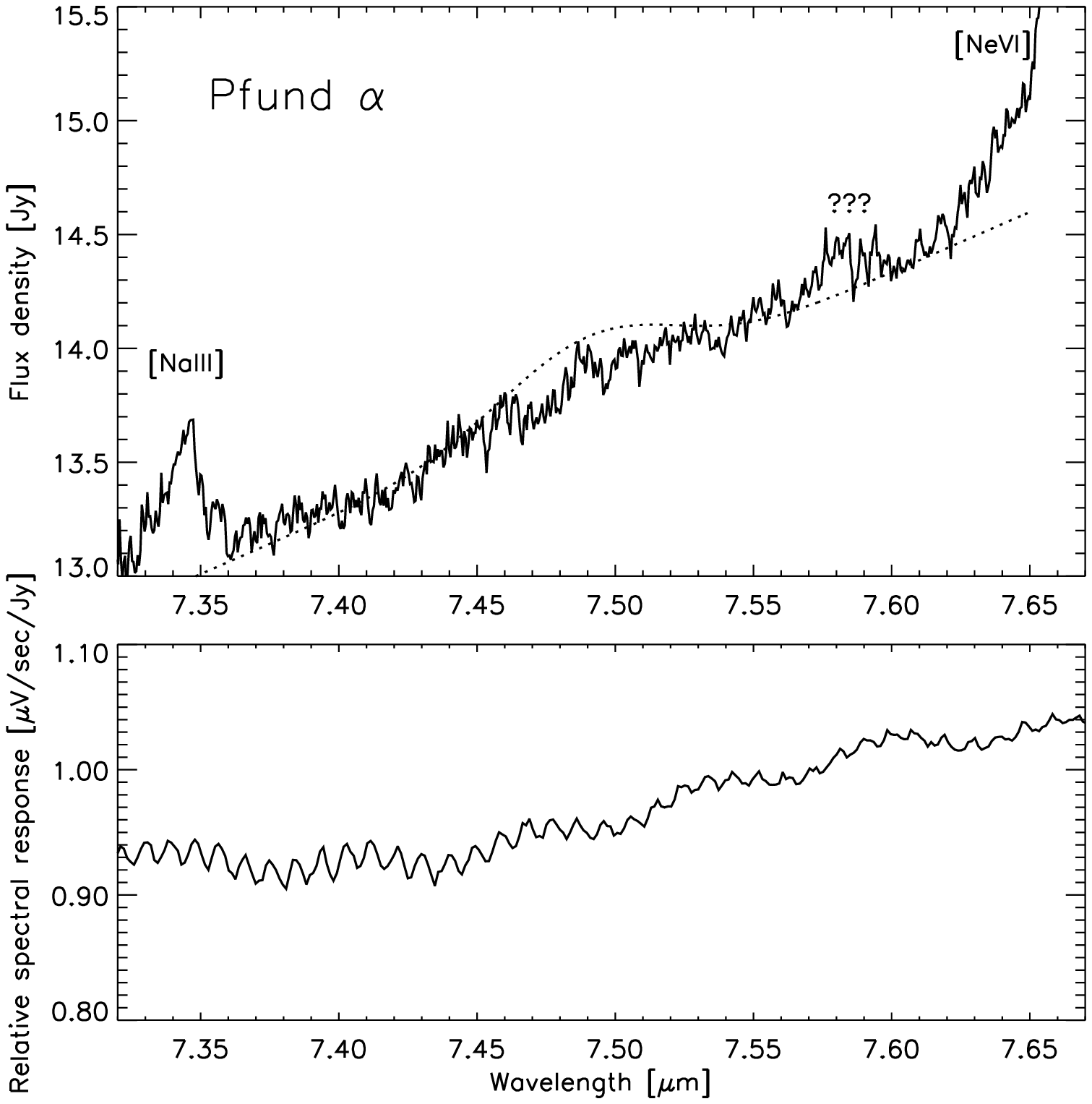]{  
As in Figure~\ref{fig:brb}, for the 7.460$\mu$m Pfund\,$\alpha$ line. 
Lines of [\ion{Na}{3}] and [\ion{Ne}{6}] are observed in this
range, as well as a possible unidentified feature. Again, the relative 
spectral response shows fringes.
\label{fig:pfa}}

\figcaption[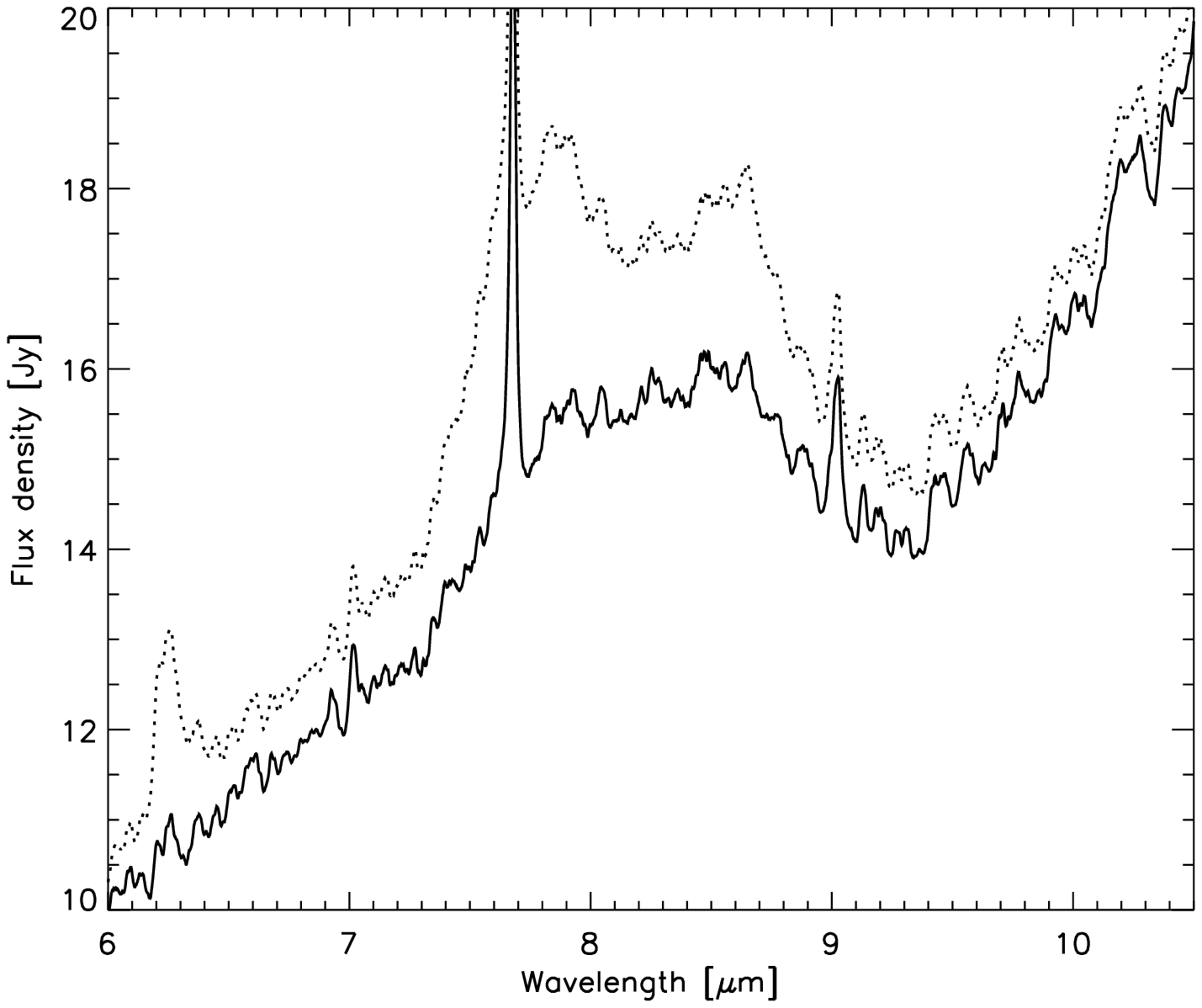]{  
The effect of PAH emission due to starburst activity on the detectability of
broad Pfund\,$\alpha$ 7.46$\mu$m emission. Part of the full SWS spectrum of 
NGC\,1068 is shown, slightly smoothed to suppress noise. The dotted line 
shows the same
spectrum, after adding a redshifted and arbitrarily 
scaled version of the spectrum of the starburst M82 intended to show the
effect of additional weak starburst activity. For clarity, emission lines 
have been omitted from the M82 spectrum. The addition of PAH
features introduces curvature to the `continuum' at the location of 
Pfund\,$\alpha$ 7.46$\mu$m. 
\label{fig:ppah}}

%
%
\clearpage

\plotone{fig_brb.eps}

\clearpage

\plotone{fig_bra.eps}

\clearpage

\plotone{fig_pfa.eps}

\clearpage

\plotone{fig_ppah.eps}


\clearpage
\begin{thebibliography}{}
\bibitem[Alexander et al. 1999]{alexander99} Alexander T., Sturm E., Lutz D., 
  Sternberg A., Netzer H., Genzel R. 1999, \apj, 512, 204
\bibitem[Antonucci \& Miller 1985]{antonucci85} Antonucci R.R.J., Miller J.S.
  1985, \apj, 297, 621
\bibitem[Bassani et al. 1999]{bassani99} Bassani L., Dadina M., Maiolino R.,
  Salvati M., Risaliti G., Della Ceca R., Matt G., Zamorani G. 1999, 
  \apjs, 121, 473
\bibitem[Blanco et al. 1990]{blanco90} Blanco P.R., Ward M.J., Wright G.S.
  1990, \mnras, 242, 4p
\bibitem[Clavel et al. 1998]{clavel98} Clavel J., et al. 1998, astro-ph/9806054
\bibitem[DePoy 1987]{depoy87} DePoy D.L. 1987, in Infrared Astronomy with
  Arrays, ed. C.G. Wynn-Williams, E.E. Becklin, \& L.H. Good (Honolulu: 
  University of Hawaii), 426 
\bibitem[Draine 1989]{draine89} Draine B.T. 1989, in: "Proc. 22nd Eslab
  Symposium on Infrared Spectroscopy in Astronomy", ESA SP-290,
  (Noordwijk: ESA), 93
\bibitem[Genzel et al. 1998]{genzel98} Genzel R., et al.
  1998, \apj, 498, 579
\bibitem[Goodrich et al. 1994]{goodrich94} Goodrich R.W., Veilleux S.,
  Hill G.J. 1994, \apj, 422, 521
\bibitem[Granato \& Danese 1994]{granato94} Granato G.L., Danese L. 1994,
  \mnras, 268, 235 
\bibitem[Granato et al. 1997]{granato97} Granato G.L., Danese L., 
  Franceschini A. 1997, \apj, 486, 147
\bibitem[de Graauw et al. 1996]{degraauw96} de Graauw Th., et al. 1996,
  \aap, 315, L49
\bibitem[Inglis et al. 1995]{inglis95} Inglis M.D., Young S., Hough J.H., 
  Gledhill T., Axon D.J., Bailey J.A., Ward M.J. 1995, \mnras, 275, 398
\bibitem[Kessler et al. 1996]{kessler96} Kessler M.F., et al. 1996, \aap, 
  315, L27
\bibitem[Koski 1978]{koski78} Koski A.T. 1978, \apj, 223, 56
\bibitem[Krolik \& Begelman 1988]{krolik88} Krolik J.H., Begelman M.C. 1988,
  \apj, 329, 702
\bibitem[Lahuis et al. 1998]{lahuis98} Lahuis F., et al. 1998, in
  Astronomical Data Analysis Software and Systems VII, A.S.P.
  Conference Series, Vol. 145, eds. R. Albrecht, R.N. Hook and H.A.
  Bushouse, p.224
\bibitem[Lutz et al. 1996]{lutz96} Lutz D.,  et al.
  1996, \aap, 315, L269
\bibitem[Lutz et al. 1997]{lutz97} Lutz D., et al. 1997, in: "First
  ISO Workshop on Analytical Spectroscopy", eds. A.M. Heras et al., ESA-SP419
  (Noordwijk:ESA), 143
\bibitem[Lutz et al. 2000]{lutz00} Lutz D., Sturm E., Genzel R., 
  Moorwood A.F.M., Alexander T., Nezter H., Sternberg A., 2000, 
  submitted to ApJ
\bibitem[Marshall et al. 1993]{marshall93} Marshall F.E., et al. 1993, \apj, 
  405, 168
\bibitem[Matt et al. 1997]{matt97} Matt G., et al. 1997, \aap, 325, L13
\bibitem[Mattila et al. 1999]{mattila99} Mattila K., Lehtinen K., Lemke D.
  1999, \aap, 342, 643
\bibitem[Miller et al. 1991]{miller91} Miller J.S., Goodrich R.W.,
  Mathews W.G. 1991, \apj, 378, 1991
\bibitem[Netzer 1990]{netzer90} Netzer H. 1990, in: Blandford R.D., Netzer H.,
   Woltjer L., "Active Galactic Nuclei", Saas-Fee advanced course 20 
   (Berlin: Springer), 57
\bibitem[Neugebauer et al. 1980]{neugebauer80} Neugebauer G., et al. 1980,
  \apj, 238, 502
\bibitem[Oliva \& Moorwood 1990]{oliva90} Oliva E., Moorwood A.F.M. 1990,
  \apj, 348, L5
\bibitem[Pier \& Krolik 1992]{pier92} Pier E.A., Krolik J.H. 1992,
  \apj, 401, 99
\bibitem[Predehl \& Schmitt 1995]{predehl95} Predehl P., Schmitt J.H.M.M. 1995,
  \aap, 293, 889
\bibitem[Rix et al. 1990]{rix90} Rix H.-W., Carleton N.P., Rieke G., Rieke M.
  1990, \apj, 363, 480
\bibitem[Ruiz et al. 1994]{ruiz94} Ruiz M., Rieke G.H., Schmidt G.D. 1994,
  \apj, 423, 608
\bibitem[Roelfsema et al. 1996]{roelfsema96} Roelfsema P.R., et al.
  1996, \aap, 315, L289
\bibitem[Shields \& Oke 1975]{shields75} Shields G.A., Oke J.B. 1975, \apj, 
  197, 5
\bibitem[Storey \& Hummer 1995]{storey95} Storey P.J., Hummer D.G. 1995,
  \mnras, 272, 41
\bibitem[Tacconi et al. 1994]{tacconi94} Tacconi L.J., Genzel R., Blietz M.,
  Cameron M., Harris A.I., Madden S. 1994, \apj, 426, L77
\bibitem[Turner et al. 1997]{turner97} Turner T.J., George I.M., Nandra K.M,
  Mushotzky R.F. 1997, \apjs, 113, 23 
\bibitem[Veilleux et al. 1997]{veilleux97} Veilleux S., Goodrich R.W., 
  Hill G.J. 1997, \apj, 477, 631
\bibitem[Verstraete et al. 1996]{verstraete96} Verstraete L., Puget J.L., 
  Falgarone E., Drapatz S., Wright C.M., Timmermann R. 1996, \aap, 315, L337
\bibitem[Ward et al. 1987]{ward87} Ward M.J., Geballe T., Smith M., 
  Wade R., Willams P. 1987, \apj, 316, 138 
\bibitem[Wieprecht et al. 1998]{wieprecht98} Wieprecht E., et al. 1998, in
  Astronomical Data Analysis Software and Systems VII, A.S.P.
  Conference Series, Vol. 145, eds. R. Albrecht, R.N. Hook and H.A.
  Bushouse, p.279
\end{thebibliography}
\end{document}